\documentclass[eqsecnum,aps,ams,preprint,epsf]{revtex4}
\newcommand{\be}{\begin{eqnarray}&&}
\usepackage{graphicx}
\usepackage{wrapfig}
\font\tenbifull=cmmib10 scaled 1200 
\font\tenbimed=cmmib9
\font\tenbismall=cmmib7
\def\bmit{\fam9 }
\mathchardef\bbgamma="710D
\mathchardef\bbkappa="7114
\mathchardef\bbrho="711A
\mathchardef\bbsigma="711B
\mathchardef\bbtau="711C
\mathchardef\bbvarrho="7125
\mathchardef\bbvarsigma="7126
\mathchardef\bbPhi="7008
\mathchardef\bbxi="7118
\def\boldgamma{{\bmit\bbgamma}}

\def\boldrho{{\bmit\bbrho}}

\def\boldtau{{\bmit\bbtau}}

\def\boldPhi{{\bmit\bbPhi}}

\def\boldeta{{\mbox{\boldmath$\eta$}}}

\newcommand{\ee}{\end{eqnarray}}

\def\dfrac{\displaystyle\frac}

\begin{document}

\thispagestyle{empty}
\title{Di-electrons from $\eta$ meson Dalitz decay
in proton-proton collisions }

\author{L.P. Kaptari$^{1,}$}
\altaffiliation{On leave of absence from
Bogoliubov Laboratory Theoretical Physics, JINR, 141980 Dubna, Russia}
\author{ B. K\"ampfer$^{1,2}$}
\affiliation{$^1$Forschungszentrum Dresden-Rossendorf, PF 510119, 01314 Dresden, Germany\\
$^2$TU Dresden, Institut f\"ur Theoretische Physik 01062 Dresden, Germany}

\date{\today}

\begin{abstract}
The reaction $ pp \to pp \eta \to pp \gamma e^+ e^-$ is discussed 
within  a covariant effective meson-nucleon theory. 
The model is adjusted to data of the subreaction $pp \to pp \eta$. 
Our focus is on
di-electrons from Dalitz decays of $\eta$  mesons, 
$\eta\to  \gamma \gamma^* \to\gamma e^+e^-$,
and the role of the corresponding transition form factor $F_{\eta \gamma \gamma^*}$.
Numerical results are presented for the intermediate energy
kinematics of HADES experiments.
\end{abstract}
\maketitle

\section{Introduction}

The $\eta$ meson as member of the octet of Goldstone bosons has the valence
quark structure $(\bar u u + \bar d d - \bar s s )/\sqrt{3}$
when choosing the mixing angle of 19.5$^o$ in superimposing the
octet-$\eta_8$ and singlet-$\eta_0$. Various conservations laws forbid
low-order decays causing the very narrow $\eta$ width. This makes the $\eta$ decays
sensitive for testing invariances of the standard model. The hidden strangeness
content lets one argue for some sensitivity to the strangeness content of the nucleon
when considering $\eta$ production off nucleons. Consequently, the $\eta$ 
production and various special decay channels were subject of intense investigations
since some time, both experimentally and theoretically. 

There is a rich data basis for $\eta$ production in nucleon-nucleon collisions
providing a test ground for meson production in strong interaction processes,
in particular near threshold. Due to the iso-scalar character of the $\eta$ meson,
$\eta$ production off the nucleon proceeds via selected baryon resonances thus
allowing differential access to resonance properties.

Furthermore, the Dalitz decay $\eta \to \gamma e^+ e^-$ constitutes a prominent source of
di-electrons in intermediate-energy heavy-ion collisions. Indeed, the recent
HADES data \cite{HADES_PRL} exhibit a sizeable yield of $e^+ e^-$ in the
invariant-mass region 150 - 500 MeV which is essentially attributed \cite{HW_HADES} to
$\eta$ Dalitz decays, but $\Delta$ Dalitz decays and non-resonant virtual bremsstrahlung
\cite{ourNuclPhys} contribute in this region, too. The primary aim of the HADES
experiments \cite{HADES} is to seek for signals of chiral symmetry restoration in compressed
nuclear matter. For such an enterprize one needs good control of the competing
background processes, among them the mentioned $\eta$ Dalitz decays.

$\eta$ Dalitz decays depend on the pseudo-scalar transition form factor.
Such form factors encode information on hadrons which is accessible in first-principle
QCD calculations or abridged variants thereof, such as effective
hadron theories or QCD sum rules. In so far, experimental information
on transition form factors is quite valuable \cite{LandPhysRep}.
Given this motivation we consider here         
the process of Dalitz decay of the pseudo-scalar $\eta$ meson
\begin{equation}
\eta \to \gamma +\gamma^* \to \gamma  + e^- + e^+,
\label{reac}
\end{equation}
where $\gamma^*$ denotes a virtual photon. 
Obviously, the probability of emitting a virtual photon is
governed by the dynamical electromagnetic
structure of the dressed transition vertex $\eta \to \gamma \gamma^*$ which
is condensed in the transition form factor $F_{\eta \gamma \gamma^*}$.
If the decaying hadron were point like, then a calculation of mass distributions
and decay widths would be straightforward along the standard quantum
electrodynamics (QED). Deviations of the measured quantities from
QED predictions directly reflect the
effects of the form factor and thus the internal hadron structure.

Often, the production process and the decay process are dealt with separately.
With respect to available new data from HADES \cite{hadeseta}
the reaction
\begin{equation}
p_1 + p_2 \to p_1' + p_2' + \eta \to p_1' + p_2' + \gamma +e^+ +e^-,
\label{dalitzpp}
\end{equation}
which will be improved in near future, we consider the complete reaction
(\ref{dalitzpp}) in which (\ref{reac}) figures as a subreaction.
The employed framework is that of an effective description in a hadronic basis.
To be specific, we are going to parameterize the $\eta$ production subreaction 
in nucleon-nucleon ($NN$) collisions
within the one-boson exchange (OBE) model. Such an approach has been utilized
fairly successfully by various authors in the past. For instance,
in Ref.~\cite{wilkinEta} a model based on
OBE with nucleons only has been facilitated for 
$pp\to pp\eta$ and $ pd\to pd\eta$ reactions.
A non-relativistic approach was proposed in Ref.~\cite{santraEta}.
In \cite{gedalinEta} a description based on OBE without internal meson conversion
and with the resonance $S_{11}(1535)$ has been elaborated for $NN\to NN\eta$ 
in a region sufficiently above the threshold.
A detailed analysis of $NN\to NN\eta$ has been worked out by
Nakayama and collaborators \cite{nakayamaEta}. 
For a study of the role of nucleon resonances in $\eta$
photo-production we refer the interested reader to Ref.~\cite{moselEta}.

In contrast to a factorized description with production of an on-shell
$\eta$ and an independent decay of an on-shell $\eta$ we attempt here
a complete description of the whole reaction (\ref{dalitzpp}).
That means we supplement the subreaction $NN \to NN \eta$ 
by the Dalitz decay part thus dealing with intermediate 
off-shell $\eta$ meson.
This allows considering di-electron masses larger
than the pole-mass of the $\eta$ meson (for the on-shell $\eta$ meson
the invariant mass of the di-electron is restricted by the
mass of the $\eta$ meson).
This can furnish additional information on the transition form factor
in a larger kinematical region.

Our paper is organized as follows. In section 2 we introduce the
$\eta \to \gamma \gamma^*$
transition form factor. Section 3 is devoted to the theoretical background for
dealing with the reaction $pp \to pp \eta \to pp \gamma e^+ e^-$.
It is essentially based on an extension of the effective model 
\cite{ourOmega,ourPhi}
adjusted to vector (V) meson  production in $NN \to NNV$ reactions. The model
utilizes a direct calculations of the relevant tree-level
Feynman diagrams within a phenomenological meson-nucleon theory.
The model parameters have been previously fixed from independent
experiments and adjusted to achieve a good description the
available experimental data \cite{ourOmega,ourPhi}. However, in the present paper
also diagrams with excitation of nucleon resonances with masses close
to the mass of a nucleon plus $\eta$ meson are included. 
These are $S_{11}(1535)$ , $D_{13}(1520)$ and $P_{11}(1440)$ resonances.
The corresponding effective constants, whenever possible, are
obtained from the known decay widths of  direct decay into $\eta$ channel or
radiative decay with subsequent use of vector meson dominance.
We use effective constants commonly adopted in the literature and
obtained from different considerations, e.g., SU(3) symmetry or adjustment  
to photoabsorbtion etc.\ \cite{nakayamaEta}.
Numerical results are presented in section 4, where
we consider separately the reaction $NN \to NN \eta$.
Results for the full reaction (\ref{dalitzpp}) are described in section 5
with emphasis on the role of the $\eta \gamma \gamma^*$ transition form factor.
The conclusions are summarized in section 6, and some
formal relations are relegated to the appendices.

\section{Dalitz decay and Transition Form Factor}

Let us first consider first the process of a two-photon decay 
of an $\eta$ meson.
The effective Lagrangian describing
the vertex $\eta \to \gamma\gamma $ reads \cite{meissner,thomas,anisovich_FF,faessler}
\begin{equation}
{\cal L}_{\eta \gamma \gamma}=f_{\eta \gamma  \gamma}\
\left( \epsilon_{\mu\nu\alpha\beta}
\partial^\mu A^\nu \partial^\alpha A^\beta
\right) \Phi_\eta,
\label{lag1}
\end{equation}
where $A^\nu$ is the electromagnetic four-potential, $\Phi_\eta$ denotes the
pseudo-scalar $\eta$ meson field and
$f_{\eta \gamma  \gamma\,}$ is the corresponding coupling constant.
The fully antisymmetric Levi-Civita symbol $\epsilon_{\mu\nu\alpha\beta}$
is normalized as $\epsilon_{0123}=-1$.
The $\eta$ decay width 
follows from (\ref{lag1}) as
\begin{eqnarray}
\Gamma_{\eta \to  \gamma \gamma}=
\frac{s_\eta^{3/2}}{64\pi}
f_{\eta\gamma \gamma}^2 
\label{g0}
\end{eqnarray}
and serves for a determination of the coupling constant $f_{\eta \gamma \gamma}$
from experimental data.
The square of the $\gamma \gamma$ invariant mass is denoted by $s_\eta$,
$s_\eta = m_\eta^2$ for an on-shell $\eta$ meson. 
(Note that contrarily to the vector meson case \cite{ourFF},
instead of a factor $1/3$ [due to averaging over three projections of the
spin of the vector particle] in eq.~(\ref{g0}) a factor of $1/2$ appears due to
two photons in the final state.)
Experimentally, the branching ratio 
$Br(\eta \to \gamma \gamma) = \Gamma_{\eta \to \gamma \gamma}/\Gamma_{\eta, tot}$ 
is known as $\left (39.38\pm 0.26\right )$\%  \cite{dataGroup}.
Eq.~(\ref{g0}) yields
$|f_{\eta  \gamma \gamma\,}|\simeq 0.025 \, GeV^{-1}$
for the known total width $\Gamma_{\eta, tot} = ( 1.3\pm 0.07) \,  keV$.
Since in our further calculations the cross section is directly proportional to
$|f_{\eta  \gamma \gamma\,}|^2$, the sign of the coupling constant
does not play a role and for definiteness it has been taken positively.

In the decay (\ref{reac}), however, one of the emitted photons 
is virtual with a time-like four-momentum and,
consequently, the Lagrangian (\ref{lag1}) must be supplemented by inclusion
of the corresponding transition form factor (FF). We employ the following procedure:
\begin{equation}
f_{ \eta \gamma \gamma}(0)\to f_{ \eta \gamma  \gamma^*}(s_{\gamma^*})
= f_{\eta \gamma  \gamma}(0) \,
F_{\eta \gamma {\gamma^*}}(s_{\gamma^*})
\label{defFF}
\end{equation}
where $s_{\gamma^*}$ is the di-electron invariant mass squared
and $f_{\eta \gamma  \gamma}(0) \equiv |f_{\eta \gamma \gamma} \vert$.
Formally, eq.~(\ref{defFF}) can be considered as the definition of the transition
form factor. By a
direct calculation of the corresponding diagram for the decay rate 
$d\Gamma / d s_{\gamma^*}$ for the $\eta$ meson one finds \cite{LandPhysRep}
\begin{eqnarray}
\dfrac{d\Gamma_{\eta\to \gamma e^+e^-}}{ds_{\gamma^*}} =
\dfrac{2\alpha}{3\pi s_{\gamma^*}}
\left(1 - \frac{s_{\gamma^*}}{m_\eta^2} \right)^3
\left(1 - \frac{4 \mu_e^2}{s_{\gamma^*}} \right)^{1/2}
\left(1 + 2 \frac{\mu_e}{s_{\gamma^*}} \right) 
\Gamma_{\eta\to  \gamma \gamma }  \left | F_{ \eta  \gamma \gamma^*\,}(s_{\gamma^*})\right |^2
\label{dgamma}
\end{eqnarray}
with $\alpha$ as electromagnetic fine structure constant
and $\mu_e$ the electron mass. In the kinematical region we are interested in
the terms with $\mu_e$ can be neglected.
Putting $\left | F_{ \eta  \gamma \gamma^*\,}(s_{\gamma^*})\right |^2 = 1$
would mean neglecting the finite size and internal structure of $\eta$.
It is seen that the differential decay width
$d\Gamma_{\eta\to \gamma e^+e^-} / ds_{\gamma^*}$
is determined by
(i) a purely kinematical (calculable) factor,
(ii) the real photon decay vertex $\eta \to \gamma \gamma$
(known from experimental data), and
(iii) the (wanted) transition FF $ F_{ \eta  \gamma \gamma^*\,}(s_{\gamma^*})  $.
Hence, eq.~(\ref{dgamma}) evidences that by measuring
the invariant mass distribution one can get direct experimental access to
the transition FF 
\cite{LandPhysRep,omegaFrmf1,omegaFrmf2,omegaFrmf3}.
For a on-mass shell $\eta$ meson, 
the value of the di-electron
mass $s_{\gamma^*}$ is kinematically restricted by
$s_{\gamma^*}\le m_\eta^2$. In case of reactions of the type (\ref{dalitzpp})
the intermediate $\eta$ meson can be off-shell and,
in principle, the value of the di-electron invariant mass can be larger than
the $\eta$ pole mass. This situation will be investigated below.

A quite successful theoretical approach to FF's is based
on the vector meson dominance (VMD) conjecture \cite{faessler,meissner}.
Reasonably good descriptions of elastic FF's in the time-like region
has been accomplished, indeed.
By using the current-field identity \cite{meissner}
\begin{equation}
J^\mu = -e\frac{M_\rho^2}{f_{\gamma\rho}}\Phi_{\rho^0}^\mu  
-e\frac{M_\omega^2}{f_{\gamma\omega}}\Phi_{\omega}^\mu 
\label{currid}
\end{equation}
with coupling constants $f_{\gamma\rho}$ and $ f_{\gamma\omega}$
known \cite{friman1,friman2}
from experimentally measured electromagnetic decay widths,
one can  also  compute the transition form
factor  $F^{VMD}_{ \eta \gamma \gamma^*\,}(s_{\gamma^*})$ by
evaluating the corresponding Feynman diagrams (see below). Contrary to the
transition FF of vector meson production \cite{ourFF}, 
the $\eta$ meson FF, computed
within such an approach, exhibits a rather good agreement with previous data
\cite{LandPhysRep,omegaFrmf1,omegaFrmf2,omegaFrmf3}.

\section{Model }
\label{subsecOdin} 

\begin{figure}[h]  
\vskip -9mm
\includegraphics[width=0.6\textwidth]{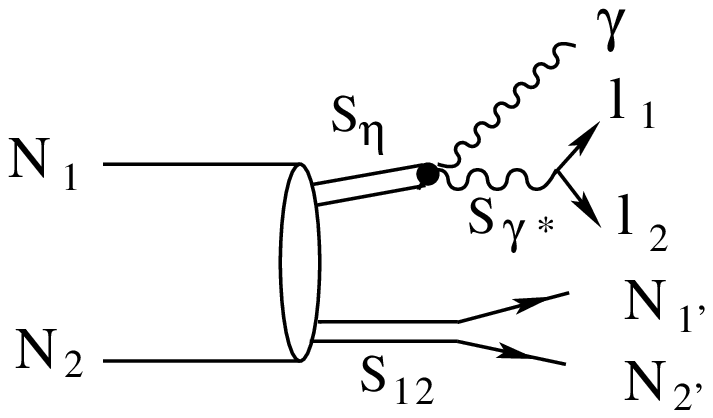} %
\caption{Illustration of the choice of the independent variables
for the process $N_1 + N_2 \to N_1' + N_2' + \gamma +l_1 + l_2$ ($l_1 = e^+$, $l_2 = e^-$) 
within the duplication
kinematics \cite{bykling}. The invariant mass squared of the two final nucleons
is denoted by $s_{12}$, while the invariant mass of the subsystem 
$\gamma e^+e^-$  is $s_\eta$. }
\label{fig1}
\end{figure}

We implement the discussed
Dalitz decay $\eta \to \gamma e^+ e^-$ into a more general process of di-electron production
in  $NN$ reactions with intermediate $\eta$. 
Consider the reaction (\ref{dalitzpp}) for which
the process (\ref{reac}) enters as a subreaction.
The invariant cross section is
\begin{eqnarray}
d^{11}\sigma =
\frac{1}{2\sqrt{\lambda(s,m_N^2,m_N^2)}}\frac{1}{(2\pi)^{11}}
\frac14 \sum\limits_{\rm spins}\
\,|\ T(P_1',P_2',k_1,k_2,k_\pi,{\rm spins}) \ |^2 d^{11}\tau_f \ \frac{1}{n! },
\label{crossnn}
\end{eqnarray}
where the factor $1/n!$ accounts for $n$ identical
particles in the final state, $\vert T \vert^2$ denotes the invariant amplitude
squared and $d\tau_f$ is the invariant phase volume which is chosen
within the so-called "duplication" kinematics \cite{bykling}, 
i.e.\ the one which exploits
invariant two-dimensional phase volumes $R_2$ describing the
decay kinematics of a  real or virtual particle with the 
invariant mass squared $s$ ($s > 0$)
into two particles, which can also be either real or virtual.
This kinematics is schematically depicted in Fig.~\ref{fig1}.
Such a choice of kinematical variables is extremely useful if one
considers specific classes of Feynman diagrams which allow to separate
some vertices in a factorized form.
Then in the total cross section some integrations can be performed analytically
(see \cite{ourNuclPhys,ourFF}). For the present task we need to
consider only such types of diagrams which
allows to factorize the meson decay vertex from
the parts describing the creation of the meson in a $NN$ interaction.
Consequently, analytical integrations can be executed over the variables
connected with the decay vertex.

\subsection{Nucleon Current}

The invariant amplitude $T$ is evaluated here
within a phenomenological meson-nucleon  theory based on
effective interaction Lagrangians which include
scalar ($\sigma$), pseudo-scalar isovector ($\pi$),
neutral pseudo-scalar ($\eta$) and
neutral vector ($\omega$) and vector isovector ($\rho$) mesons (see
\cite{nakayamaEta,ourPhi,ourOmega,ourNuclPhys})
\begin{eqnarray}
{\cal L}_{\sigma NN }&=& g_{\sigma NN} \bar N  N \it\Phi_\sigma , \label{mnn1}\\
{\cal L}_{\pi NN}&=&
-\frac{f_{ \pi NN}}{m_\pi}\bar N\gamma_5\gamma^\mu \partial_\mu
({\boldtau \boldPhi_\pi})N ,\\
{\cal L}_{\eta NN}&=&
-\frac{f_{\eta NN}}{m_\eta}\bar N\gamma_5\gamma^\mu \partial_\mu
\Phi_\eta N , \\
{\cal L}_{\rho NN}&=&
-g_{\rho NN }\left(\bar N \gamma_\mu{\boldtau}N{\boldPhi_ \rho}^\mu-\frac{\kappa_\rho}{2m_N}
\bar N\sigma_{\mu\nu}{\boldtau}N\partial^\nu{\boldPhi_\rho}^\mu\right) , \\
{\cal L}_{\omega NN}&=&
-g_{\omega  NN }\left(
\bar N \gamma_\mu N {\it\Phi}_{\omega}^\mu-
\frac{\kappa_{\omega}}{2m_N}
\bar N \sigma_{\mu\nu}  N \partial^\nu \it\Phi_{\omega}^\mu\right) ,
\label{mnn}
\end{eqnarray}
where $N$ and $\it\Phi$ denote the nucleon and meson fields, respectively
and bold face letters stand for isovectors.
All couplings with off-mass shell
mesons are dressed by monopole form factors
$F_M=\left(\Lambda^2_M-\mu_M^2\right)/\left(\Lambda^2_M-k^2_M\right)$,
where $k^2_M$ is the four-momentum of a virtual meson with mass $\mu_M$.

\begin{figure}[h]  
\vskip 3mm
\includegraphics[width=1.0\textwidth]{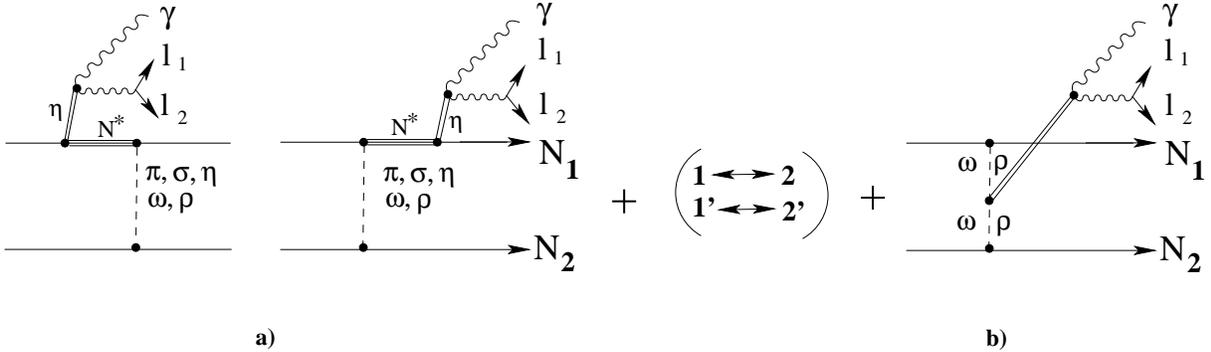} %
\vskip -3mm
\caption{Diagrams for the process
$N N \to N N  \gamma e^+e^-$ within an effective meson-nucleon theory.
a) Dalitz decays of $\eta$ mesons from bremsstrahlung like diagrams.
The intermediate
baryon $N^*$ (triple line) can be either a nucleon or one of the  nucleon resonances
$S_{11}(1535)$, $P_{11}(1440)$ or $D_{13}(1520)$. Analog diagrams for the
emission from Fermion line $N_2$.
b) Dalitz decay of $\eta$ mesons from internal meson conversion.
Exchange diagrams for identical nucleons in exit channel are not displayed.}
\label{fig2}
\end{figure}

The Lagrangians (\ref{mnn1} - \ref{mnn}) are needed in evaluations of
Feynman diagrams describing
the Dalitz decay of the $\eta$ meson created from nucleon
bremsstrahlung due to $NN$ interaction via 
one-boson exchange, see Fig.~\ref{fig2}a.

\subsection{Internal Conversion Current}

The  $\eta$ meson can also be produced by an internal conversion
of the exchanged mesons, the so-called conversion current.
The dominant exchange vector mesons ($V$) in this case
are $\omega$ and
$\rho$ mesons with the interaction Lagrangians
\begin{eqnarray}
{\cal L}_{\eta \omega\omega}&=& -\frac{g_{\eta \omega\omega}}{2m_\omega}\,
\varepsilon_{\mu\nu\alpha\beta}
\,\left (\partial^\mu\it \Phi_\omega^\nu \partial^\alpha\it
\Phi_\omega^\beta \right)\Phi_\eta, \\
{\cal L}_{\eta \rho\rho}&=& -\frac{g_{\eta \rho\rho}}{2m_ \rho}\,
\varepsilon_{\mu\nu\alpha\beta}
\,\left (\partial^\mu\it \boldPhi_ \rho^\nu    \partial^\alpha\it
\boldPhi_ \rho^\beta \right)\Phi_\eta.
\label{conversion}
\end{eqnarray}
The corresponding diagrams are exhibited in Fig.~\ref{fig2}b.

The $VV\eta$ vertices in the conversion diagrams have been calculated from the
radiative decay $V\to \eta\gamma$ within the VMD model \cite{nakayamaEta,durso}.
Correspondingly, the vertex form factor $VV\eta$ is chosen as
\begin{equation}
\label{convFF}
F_{VV\eta}(\Lambda,k_1^2,k_2^2)=
\frac{\Lambda^2-m_V^2}{\Lambda^2-k_1^2}\frac{\Lambda^2}{\Lambda^2-k_2^2}
\end{equation}
which, in accordance with the procedure of determining the
coupling constant,
is normalized to unity when one vector meson is on-mass shell and the other one
becomes massless, e.g., $F_{VV\eta}(\Lambda,k_1^2=m_V^2,k_2^2=0)=1$.

\subsection{Nucleon Resonance Current}

In the threshold-near kinematics for $\eta$ production in $NN$
reactions there are a few nucleon resonances with masses
$m_{N^*} \sim m_N + m_\eta$ which can contribute to the cross section.
These are $S_{11}(1535)$ and $D_{13}(1520)$ with odd parity and spins
$\frac12$ and $\frac 32$, respectively, and the spin-$\frac12$ even-parity
$P_{11}(1440)$ Roper resonance with nucleon quantum numbers. 
The corresponding interaction Lagrangians 
for $S_{11}(1535)$ are \cite{nimaiResonances,nakayamaEta}
\begin{eqnarray}
{\cal L}_{\eta  NN_{1535} }&=& \frac{g_{\eta   NN_{1535}}}{m_{N^*}-m_N}
 \bar N^*\gamma_\mu \partial^\mu  \Phi_\eta N +h.c. ,\label{res121}\\
{\cal L}_{\pi  NN_{1535} }&=& \frac{g_{\pi   NN_{1535}}}{m_{N^*}-m_N}
 \bar N^*\gamma_\mu \partial^\mu \boldtau \boldPhi_\pi N +h.c., \\
{\cal L}_{\omega  NN_{1535} }&=& \frac{g_{\omega  NN_{1535}}}{m_{N^*}+m_N}
\bar N^* \gamma_5\sigma_{\mu\nu} \partial^\nu\Phi^\mu_\omega N +
h.c., \\
{\cal L}_{\rho  NN_{1535} }&=& \frac{g_{\rho  NN_{1535}}}{m_{N^*}+m_N}
\bar N^* \gamma_5\sigma_{\mu\nu} \partial^\nu \boldtau\boldPhi^\mu_\rho N +
h.c. .
\label{res12}
\end{eqnarray}

For the spin-$\frac32$ resonance $D_{13}(1520)$ we employ
\begin{eqnarray}
{\cal L}_{\eta  NN_{1520} }&=& \frac{g_{\eta   NN_{1520}}}{m_\eta}
 \bar N^{*\mu} \Theta_{\mu\nu}(z) \gamma_5 \partial^\nu  \Phi_\eta N +h.c.,\label{res321}\\
{\cal L}_{\pi  NN_{1520} }&=& \frac{g_{\pi   NN_{1520}}}{m_\pi}
 \bar N^{*\mu} \Theta_{\mu\nu}(z) \gamma_5 \partial^\nu \boldtau
\boldPhi_\pi N +h.c. ,\\
{\cal L}_{\omega  NN_{1520} } &=&
i\frac{g^{(1)}_{\omega   NN_{1520}}}{2m_N}
\bar N^{*\mu} \Theta_{\mu\nu}(z)\gamma_\alpha \omega^{\alpha\nu} N \, -\,
\frac{g^{(2)}_{\omega   NN_{1520}}}{4m_N^2}
\partial_\alpha\bar N^{*\mu}  \omega^{\alpha\mu} N + h.c.,\\
{\cal L}_{\rho  NN_{1520} }&=&
i\frac{g^{(1)}_{\rho   NN_{1520}}}{2m_N}
\bar N^{*\mu}\Theta_{\mu\nu}(z) \gamma_\alpha \boldtau\boldrho^{\alpha\nu} N
\, -\,\frac{g^{(2)}_{\rho   NN_{1520}}}{4m_N^2}
\partial_\alpha\bar N^{*\mu}  \boldtau\boldrho^{\alpha\nu} N + h.c.,
\label{res32}
\end{eqnarray}
where $\Theta_{\mu\nu}(z)=g_{\mu\nu} -(A(1+4z)/2+z)\gamma_\mu\gamma_\nu$.
The field strengths
$\omega^{\alpha\mu}$ and $\boldrho^{\alpha\nu}$ are
$\omega^{\alpha\mu}(x) = \partial^\alpha \Phi_\omega^\mu(x)-
\partial^\mu\Phi_\omega^\alpha(x)$
and
$\boldrho^{\alpha\nu}(x) = \partial^\alpha \boldPhi_\rho^\mu(x)-
\partial^\mu\boldPhi_\rho^\alpha(x)$,
respectively. The interactions (\ref{res321} - \ref{res32}) with
$\Theta_{\mu\nu}(z)$ have been extensively
discussed in the literature (see, e.g., Ref.~\cite{davitsonDelta} and
further references quoted therein).
The form of $\Theta_{\mu\nu}(z)$ is chosen in such a way that the
respective interaction Lagrangian
obeys the same point transformation as the free Lagrangian, which
ensures that the matrix elements are independent
of the parameter $A$, usually taken as $A=-1$. The off-shell parameter $z$
remains free thereby.
We employ here $z=-1/2$. Also, the choice of the form of higher
spin $\bar s \ge \frac32$ propagators has been a subject of discussion in the
literature \cite{ourNuclPhys,prop4,prop5,prop6,prop7}, e.g., with concerns
about the spin-projector operator ${\cal P}_{\frac32}$ (off-mass shell vs.\
on-shell) or the ordering of the product of energy projection
operator $\hat P_{N^*}+m_{N^*}$ and ${\cal P}_{\frac32}$
(only on the mass shell these two operators commute). In the present paper we
take the spin-$\frac32$ propagator in the form \cite{ourNuclPhys}
\begin{eqnarray}
\label{prop32}
S_{\frac32}^{\mu\nu}(P)&=&-i\frac{\hat P_{N^*}+m_{N^*}}{P^2-m_{N^*}^2}
{\cal P}_{\frac32}^{\mu\nu} (P),
\end{eqnarray}
where the spin projection operator is of the form 
\begin{eqnarray}
\label{spr32}
{\cal P}_{\frac32}^{\mu\nu} (P) &=&  g^{\mu\nu}-\frac13\gamma^\mu\gamma^\nu
-\frac{2P^\mu P^\nu}{3m_{N^*}^2}
-\frac{1}{3m_{N^*}}\left( \gamma^\mu P^\nu-\gamma^\nu P^\mu\right),
\end{eqnarray}
as commonly adopted within the Rarita-Schwinger formalism \cite{fron}.

For $P_{11}(1440)$ 
the effective Lagrangians are similar to that for $S_{11}$,
except for the relative signs in $\eta $ and $\pi$ couplings
and the $\sigma NN^*$ Lagrangian, i.e.
\begin{eqnarray}
{\cal L}_{\eta  NN_{1440} }&=& -\frac{g_{\eta   NN_{1440}}}{m_{N^*}+m_N}
 \bar N^*\gamma_5\gamma_\mu \partial^\mu  \Phi_\eta N +h.c.,\\
{\cal L}_{\pi  NN_{1440} }&=&- \frac{g_{\pi   NN_{1535}}}{m_{N^*}+m_N}
 \bar N^*\gamma_5\gamma_\mu \partial^\mu \boldtau \boldPhi_\pi N +
h.c.,\\
 {\cal L}_{\sigma NN_{1440} }&=& g_{\sigma NN_{1440}} \bar N^*  N \it\Phi_\sigma
+h.c., \\
{\cal L}_{\omega  NN_{1440} }&=& \frac{g_{\omega  NN_{1440}}}{m_{N^*}+m_N}
\bar N^*  \sigma_{\mu\nu} \partial^\nu\Phi^\mu_\omega N +h.c., \\
{\cal L}_{\rho  NN_{1440} }&=& \frac{g_{\rho  NN_{1440}}}{m_{N^*}+m_N}
\bar N^*  \sigma_{\mu\nu} \partial^\nu \boldtau\boldPhi^\mu_\rho N +
h.c.   .
\label{res14}
\end{eqnarray}
These interaction Lagrangians enter the nucleon resonance current
exhibited in Fig.~\ref{fig2}a.

In order to account for the finiteness of the resonance widths $\Gamma$,
the resonance mass in the corresponding  propagators is augmented by
an imaginary part, $m_{N^*}\to  m_{N^*} -i\Gamma_{N^*}/2$.
Also for the resonance and conversion currents 
all the vertices with off-shell hadrons
are dressed by form factors
\begin{equation}
\label{ffversh}
F(\Lambda,p^2)=\frac{\Lambda^4}{ \Lambda^4 +(p^2-m^2)^2  }.
\end{equation}

\subsection{Effective parameters}

The effective parameters of the Lagrangians (\ref{mnn1} - \ref{mnn}) which determine
the pure nucleon current diagrams are basically
the ones from the OBE Bonn potential \cite{bonncd} which have been used in
our previous calculations for vector meson ($\omega$, $\rho$ and $\phi$)  
production in $NN$ interactions \cite{ourOmega,ourFF}, 
see Tab.~\ref{tb1}. The nucleon cut-off
parameter for the bremsstrahlung $NN\eta$ vertex is 
$\Lambda^{br.}_{\eta N N}=1.2 \,GeV$.

\begin{table}[!ht]
\caption{\it Coupling constants and cut-off
parameters for the nucleon current \cite{bonncd}.}
\begin{tabular}{l c l c c }\hline\hline 
\multicolumn{5}{c}{}\\[-6mm]
                   &\phantom{ppp} &   $g_{MNN} $          & &   $\Lambda_{MNN} \,[GeV]$ \\ \hline
    $\pi $                & & $f_{ \pi NN}=1.0$                & & $ 1.3$ \\
    $\eta$                & & $f_{ \eta NN}=1.79$              & & $ 1.8$ \\
    $\sigma$              & & $g_{\sigma NN}=10.$              & & $ 1.8$ \\
    $\rho$                & & $g_{\rho NN}=3.5$                & &$1.3$   \\
                          & & ($\kappa_\rho=6.1$)              & &        \\
    $\omega$              & &  $g_{\omega NN}=15.85$           & &$1.5$   \\
                          & & ($\kappa_\omega=0.0$)            & &        \\
  \hline\hline
\end{tabular}
\label{tb1}
\end{table}

The effective coupling constants for the nucleon resonance currents, 
whenever possible, have been
obtained from the known decay widths of  direct decay into $\eta$ channels or
radiative decay with subsequent use of VMD.
Otherwise, we use effective constants commonly utilized in the literature and
obtained from different considerations, e.g., SU(3) symmetry, fit of photo-absorbtion
reactions etc.\ (see, e.g.\ \cite{nakayamaEta}). The few remaining less known cut-off
parameters are taken either close to the ones from OBE potential
(for instance, the $\Lambda_{\rho N N^*}$ and $\Lambda_{\omega N N^*}$ cut off's are
chosen equal to $\Lambda_{\omega NN}=1.5\, GeV$;
$\Lambda_{\eta NN_{1535}}=\Lambda_{\eta NN_{1440}}=\Lambda_{\eta NN}=1.3\, GeV$,
cf.\ Tab.~\ref{tb2})
or adjusted to experimental data (see below). The resonance cut-off
parameters for the bremsstrahlung $N^*N\eta$ vertex have been taken as 
$\Lambda^{br.}_{\eta N N_{1535}}=1.3 \,GeV$,
$\Lambda^{br.}_{\eta N N_{1520}}=1.1 \,GeV$ and  
$\Lambda^{br.}_{\eta N N_{1440}}=1.2 \,GeV$,
respectively.

\begin{table}[!ht]
\caption{\it Coupling constants and cut-off 
parameters for the nucleon resonance current. 
For the spin-$\frac32$ resonance
$D_{13}$ the off-shell parameter $z=-1/2$ is used and 
the second coupling constant, eq.~(\ref{res32}), with
vector mesons is given in parenthesis.
For a reasoning of $g_{\pi NN^*}$ and $g_{\eta NN^*}$ for
$N^* = D_{13}(1520)$ see Appendix A.}
\begin{tabular}{cc  cccc cccc cccc }\hline\hline
\multicolumn{2}{c}{\phantom{vert}}
  &\multicolumn{4}{c }{$S_{11}(1535)$}
  & \multicolumn{4}{ c }{$D_{13}(1520)$}
  &\multicolumn{4}{c }{$P_{11}(1470)$} \\
  &  && $g_{MNN^*}$         &&$\Lambda \,[GeV]$ &&$g_{MNN^*}$ & &$\Lambda \, [GeV]$ &&$g_{MNN^*}$ & &$\Lambda \,[GeV]$ \\\hline
  $\pi$&&&    1.25 & &1.2     &&  1.55   &&1.0 &&6.54&&1.3\\
 $\eta$&&&    2.02 & &1.3     &&  8.3   &&1.2 &&0.49 &&1.3\\
 $\omega$&&& -0.72&  &1.5     &&  -2.1(0.7)   &&1.5 &&-0.37&&1.5\\
 $\rho$&&& -0.65 &   &1.5     &&  6(-2.1)   &&1.5 &&-0.57 &&1.5\\ \hline\hline
\end{tabular}
\label{tb2}
\end{table}

The coupling constants $g_{\omega\omega\eta}$ and
$g_{\rho\rho\eta}$ for the conversion current have been calculated \cite{nakayamaEta}
from a combined analysis of
the radiative decay within the VMD model and with a SU(3) effective
Lagrangian which provides
\begin{equation}
\label{omom}
g_{\omega\omega\eta} \simeq 4.85; \quad g_{\rho\rho\eta}\simeq 4.95.
\end{equation}
(Note that a naive direct calculations of these constants within the VMD model
can provide slightly larger values $g_{VV\eta}\sim 6.0$.) The corresponding
cut-offs are $\Lambda_{\omega\omega\eta}=\Lambda_{\rho\rho\eta}=1.6 \,GeV$.

\subsection{Cross section for  $\bf pp \to pp \boldeta$} 

The invariant amplitude $T$ in eq.~(\ref{crossnn})
corresponding to the Feynman diagrams in Fig.~\ref{fig2} 
can be cast in a factorized form
\begin{eqnarray}
\label{ampl}
T=T^{(1)}_{NN\to NN\eta}\dfrac{i}{q^2-(m_\eta -\frac{i}{2} \Gamma_{\eta,tot})^2}\,
T^{(2)}_{\eta\to\gamma e^+e^-},
\end{eqnarray}
where the amplitude $T^{(1)}_{NN\to NN\eta}$
describes the process of production of an off-shell
$\eta$ meson in a $NN$ collision, while the amplitude
$T^{(2)}_{\eta\to\gamma e^+e^-}$ describes the Dalitz decay of the produced
$\eta$ meson into a real photon and a di-electron. In the propagator of the $\eta$ meson
the mass $m_\eta$ has been replaced by $m_\eta - i\Gamma_{\eta,tot}/2$ to take into account
the finite life time of the $\eta$ meson.

As mentioned, for such factorized Feynman diagrams one can separate, in the cross section,
the dependence on the variables connected with the Dalitz decay vertex and
perform the phase space integration analytically.
Equation (\ref{ampl}) allows one to rewrite the differential cross section 
(\ref{crossnn}) in a factorized form as well,
\begin{eqnarray}
\frac{d\sigma}{ds_\eta ds_{\gamma^*}} &=&
\frac{2 \alpha}{3 \pi^2} \, \vert F_{\eta \gamma \gamma^*} \vert^2 \,
\Gamma_{\eta \to \gamma \gamma}
\frac{(s_\eta - s_{\gamma^*})^3}{m_\eta^3 s_\eta s_{\gamma^*}}
\frac{1}{(s_\eta - m_\eta^2)^2 + m_\eta^2 \Gamma_{\eta, tot}^2} 
\, d^5\sigma^{tot}_{NN\to NN\eta} \\
&\approx&\dfrac{d\Gamma_{\eta\to \gamma e^+e^-}}{ds_{\gamma^*}}\,
\frac{1}{4\pi}\frac{ s_\eta^{-1/2}}{(\sqrt{s_\eta}-m_\eta)^2 +\frac14 \Gamma_{\eta, tot}^2}
\, d^5\sigma^{tot}_{NN\to NN\eta},
\label{two}
\end{eqnarray}
where the integral over the final di-electron and
photon variables has been performed analytically 
(see for details \cite{ourFF,ourNuclPhys}). 
$d\Gamma_{\eta\to \gamma e^+e^-} / ds_{\gamma^*}$ is defined by 
eq.~(\ref{dgamma}) but with $m_\eta \to s_\eta$
and the production cross section of a pseudo-scalar meson with
$\eta$ quantum numbers but with ${s_\eta}\neq m_\eta^2$ is
\begin{eqnarray}
\label{nneta}
d^5\sigma^{tot}_{NN\to NN\eta}
&=&\frac{1}{2(2\pi)^5\sqrt{\lambda(s,m_N^2,m_N^2)}} \nonumber\\
&\times& \frac14
\sum_{spins}| T_{NN\to NN\eta}^{(1)}|^2 ds_{12}dR_2(N_1N_2\to s_\eta s_{12}) 
dR_2 (s_{12}\to N_1'N_2'),
\end{eqnarray}
where the two-particle invariant phase space volume $R_2$ reads
\begin{equation}
R_2(ab\to cd)=\frac{\sqrt{\lambda(s_{ab},m_c^2,m_d^2)}}{8s_{ab}} d\Omega^*_c .
\label{ph}
\end{equation}
In what follows
we are interested in production and subsequent decay of intermediate $\eta$ mesons, 
thus being off-mass shell, $s_\eta\neq m_\eta^2$.
We assume that for the off-shell
$\eta$ the net coupling constant
$f_{\eta\gamma\gamma}(0)$ in eq.~(\ref{defFF}) obtained from the experimental data
via eq.~(\ref{dgamma}) remains the same.
The off-mass shellness of $\eta$ mesons is taken into account by including
additional effective form factors in the corresponding vertices.

\section{$\boldeta$ Production Cross Section}

It can be seen from (\ref{two}) that the peculiarities of the
cross section for the full reaction $NN\to NN \gamma e^+e^-$
are basically determined by the subreaction $NN\to NN\eta$
with creation of a virtual $\eta$ meson. Hence, before
analyzing the full reaction, we proceed with a study of
the subreaction $NN\to NN\eta$ for production of an on-shell $\eta$ meson.

\subsection{Initial and Final State Interactions}

It ought to be mentioned that the Feynman diagrams
depicted in Fig.~\ref{fig2} cover only the process of creation and decay of
the pseudo-scalar $\eta$ meson, the so-called
production current. However, in the complete process (\ref{dalitzpp}) the two nucleons
can suffer initial state interaction (ISI) 
and final state interaction (FSI), before and after the $\eta$ creation,  
thus provoking distortions of the incoming and outgoing $N N$ waves.

The ISI within a $NN$ pair before the $\eta$ creation
is to be evaluated at relatively high energies, larger than the
threshold of the $\eta$ meson production ($T_{kin}\sim 1.3 \,GeV$).
Therefore, one can expect that the variation of ISI effects 
with the kinetic energy is small.
Indeed, as shown in Ref.~\cite{isi}, the effect of ISI
can be factorized in the total cross section and it plays effectively the role of a
reduction factor in each partial wave in the cross section. This reduction factor
depends on the inelasticity and  phase shifts of the
partial waves at the considered energies.
Near the threshold, the number of initial partial waves is strongly limited
by the partial waves of the final states, and
one can restrict oneself to $^3P_0$ and $^1P_1$ waves.
Experimentally \cite{said} it is found that at kinetic energies of the order of few $GeV$
the phase shifts $^3P_0$ and $^1P_1$ are indeed almost energy independent and the
reduction factor for each partial wave can be taken as constant.
In our calculations we adopt for the reduction factor $\zeta$ the expression
from Ref.~\cite{isi}
\begin{eqnarray}
\label{isi}
\zeta_i= \varsigma_i(p)\cos^2 \delta_i(p)+\frac14\left( 1-\varsigma_i(p)\right)^2,
\end{eqnarray}
where $\varsigma_i$ and $\delta_i$ denote the inelasticity and phase shifts,
respectively. We employ in our calculations
$\zeta=0.277$ for the $^1P_1$ wave ($i = 1$) and
$\zeta=0.243$ for $^3P_0$ ($i = 2$) \cite{nakayamaEta}.

FSI effects among the escaping $NN$ pair are accounted for within
the Jost function formalism \cite{gillespe}
which reproduces the singlet and triplet phase shifts at low energies,
as appropriate for reactions near threshold. 
Details of calculations of FSI with the Jost function can be found in Ref.~\cite{fsiReznik}.

\subsection{$\boldeta$ Meson Production in $\bf NN$ Collisions}

\begin{figure}[h]  
\includegraphics[width=1.0\textwidth]{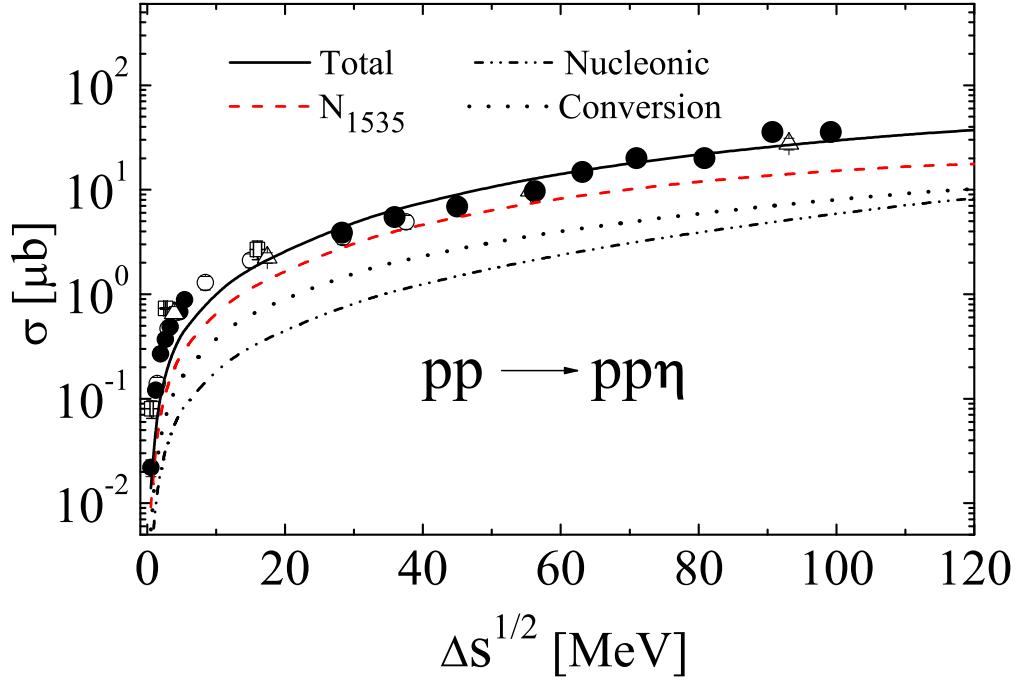}
\vskip -9mm
\caption{Total cross section for the reaction $pp\to pp\eta$ 
as a function of the excess energy $\Delta s^{1/2} = \sqrt{s}-2m-m_\eta$.
The dot-dashed line corresponds to the nucleon
current contribution (without resonances, Fig.~\ref{fig2}a),
while the dotted line is for the
internal meson conversion diagram (Fig.~\ref{fig2}b).
The dashed curve exhibits the contribution
of the $S_{11}(1535)$ resonance (nucleon resonance current 
in Fig.~\ref{fig2}a). Contributions from
$P_{11}(1440)$ and $D_{13}(1520)$ are smaller and not displayed.
The solid curve is the total contribution with all interferences.
Data are from Refs.~\cite{{etacalen},etaexper,etaangular,etaexper1,etaexper2,etaexper3};
error bars are suppressed.}
\label{fig3}
\end{figure}

\begin{figure}[h]  
\includegraphics[width=1.0\textwidth]{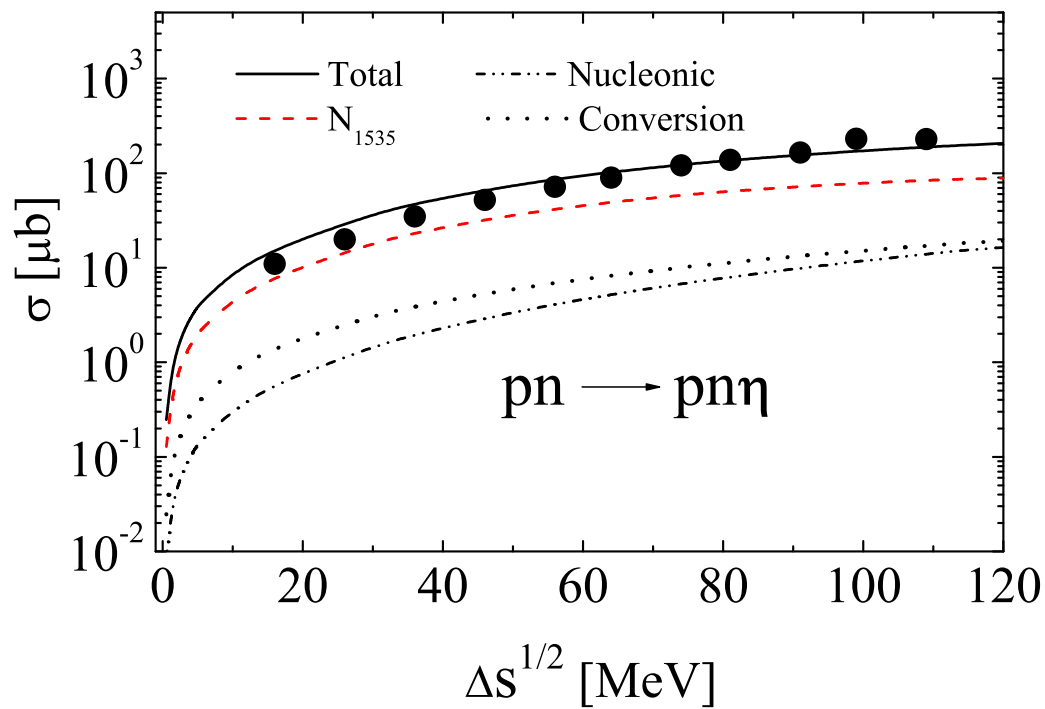}
\vskip -9mm
\caption{The same as in Fig.~\ref{fig3} but for the reaction $pn\to pn\eta$.}
\label{fig4}
\end{figure}

Results for the energy dependence  of the total cross section
$\sigma^{NN\to NN\eta}$, eq.~(\ref{nneta}), are presented in 
Figs.~\ref{fig3} and \ref{fig4},
for proton-proton and proton-neutron reactions, respectively.
The amplitude $T_{NN\to NN\eta}^{(1)}$ represents a sum of
nucleon, meson conversion and nucleon resonance currents, 
each of them being a sum over all the
considered exchanged mesons (see Fig.~\ref{fig2}).
Using the parameters listed in Tabs.~\ref{tb1} and \ref{tb2} 
the nucleon current contribution
with parameters from the Bonn group \cite{bonncd}
is found to be fairly small in the present calculations (see dot-dashed curves).

Also the contribution from the conversion
current (see doted curves) is not too large. 
The main contribution to the cross section
comes from the nucleon resonance currents. 
Here it is worth mentioning that, in spite of the
large number of the considered diagrams and 
the  large number of the effective
parameters, there is not too much freedom for fine-tuning of the cross section. 
As mentioned, most parameters are restricted  by independent experiments and they cannot
be varied in large intervals. We can slightly modify the less known parameters to achieve
improvement of the overall description with data. 
In particular, in the present work we find a small contribution of
the resonance $D_{13}(1520)$ at small energies, but a rather strong energy dependence
due to $\pi$ and $\eta$ exchange diagrams with increasing energy.
Therefore, in order to reduce the influence of the $D_{13}(1520)$
at large excess energies, for these diagrams the cut-off parameters
have been chosen smaller than for others (see Tab.~\ref{tb2}).

Note that, even achieving a good description of the cross section in proton-proton
reactions, it is not \textit{a priory}
clear that the obtained set of parameters equally well describes also the proton-neutron
reactions. The isospin dependence of the amplitude is determined by a subtle
interplay of different diagrams with different exchange mesons (scalar, vector,
isoscalar, isovector). Once the parameters for the $pp$ reaction
are fixed, the $pn$ amplitude follows directly  without further parameters
(ISI and FSI are different for $pp$ and $pn$ systems, but fixed independently). 
Figure \ref{fig4} demonstrates that the isospin dependence of the
amplitude is correctly described with respect to available data. 

\begin{figure}[h]  
\vskip -9mm
\includegraphics[width=1.0\textwidth]{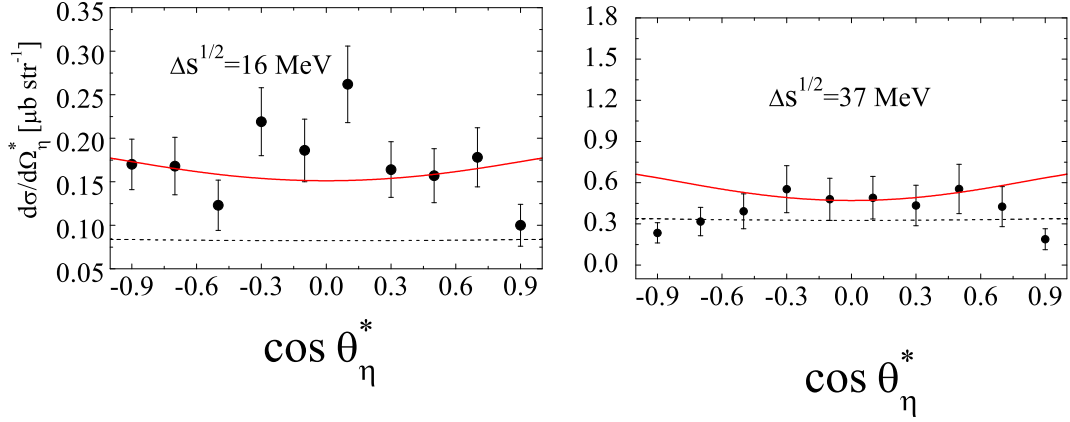} %
\vskip -9mm
\caption{Angular distributions of $\eta$ mesons in the $pp$ center of
mass system for the excess energies $\Delta s^{1/2} = 16 \,MeV$
(left panel) and $37 \, MeV$ (right panel). The dotted lines depict the contribution
of the conversion currents, while the solid lines correspond to the
total differential cross section, including also nucleon and resonance 
currents and interferences. Data are from Ref.~\cite{etaangular}.}
\label{fig5}
\end{figure}

In Fig.~\ref{fig5} the angular distribution
of $\eta$ mesons in the center of mass is presented for two excess energies,
$\Delta s^{1/2}=16 \,MeV$ (left panel) and $ \Delta s^{1/2}=37 \,MeV$ (right panel).
Here a few comments are in order. 
Generally, at threshold-near energies the angular distributions
are rather flat. That means that the total
cross section at the corresponding energy is 
$\sigma^{tot} \propto 4\pi d\sigma/d\Omega^*$ and
a quantitatively good description of the angular distribution
$d\sigma/d\Omega^*$ provides, of course, also a good description
of the total cross section. Therefore we compare our calculations with data at such 
an energy, where we achieve a better description of the total cross section
(cf.\ Fig.~\ref{fig3}).  The other existing experimental data at similar
energies (in particular, $\Delta s^{1/2}=15.5 \, MeV$, \cite{etaangular}) 
differ from each other
by an overall normalization factor, which can amount up to $40\%$ at these energies,
see Ref.~\cite{etacalen}. Another important issue for the analysis
of the angular distributions is the shape at forward and backward directions.
The conversion current provides an upward ($\bigcup$) shape of the distributions, while the
nucleon and nucleon-resonance currents generally give a downward ($\bigcap$) curve at
forward and backward directions. Since in our calculations the contribution 
of the conversion current is much smaller than the one from the nucleon resonance current,
the resulting angular distribution,
albeit being rather flat, has however an upward ($\bigcup$) trend.

In Fig.~\ref{fig6} the invariant mass distribution of the two final protons 
is exhibited together
with experimental data at an excess energy of $\Delta s^{1/2}=15.5\, MeV$. Since our total
cross section has been optimized to describe the data at higher energies, 
$\Delta s^{1/2}>16\,MeV$,
see Fig.~\ref{fig3}, the calculated curve in Fig.~\ref{fig6} 
has been multiplied by a factor
of $\approx 1.4$ obtained from an attempt to reconcile the data
at $\Delta s^{1/2}=15.5\, MeV$ and $\Delta s^{1/2}=16 \,MeV$
(relevant comments and discussion about the normalization of data can be
found in Refs.~\cite{etaangular,etacalen}).

The mild discrepancy of the angular distribution in Fig.~\ref{fig5}
and the slight underestimate of the total cross section data at
$\Delta s^{1/2} \sim 10 \, MeV$ let us argue that further reaction 
details could be accommodated. For instance, the $\bar s s$ shake-off
considered in \cite{TitovBK} may be included. This, however, deserves
a separate consideration. As interim summary we believe to have at our disposal
an appropriate description of the $\eta$ production cross section.

\begin{figure}[h]  
\vskip -13mm
\includegraphics[width=1.0\textwidth]{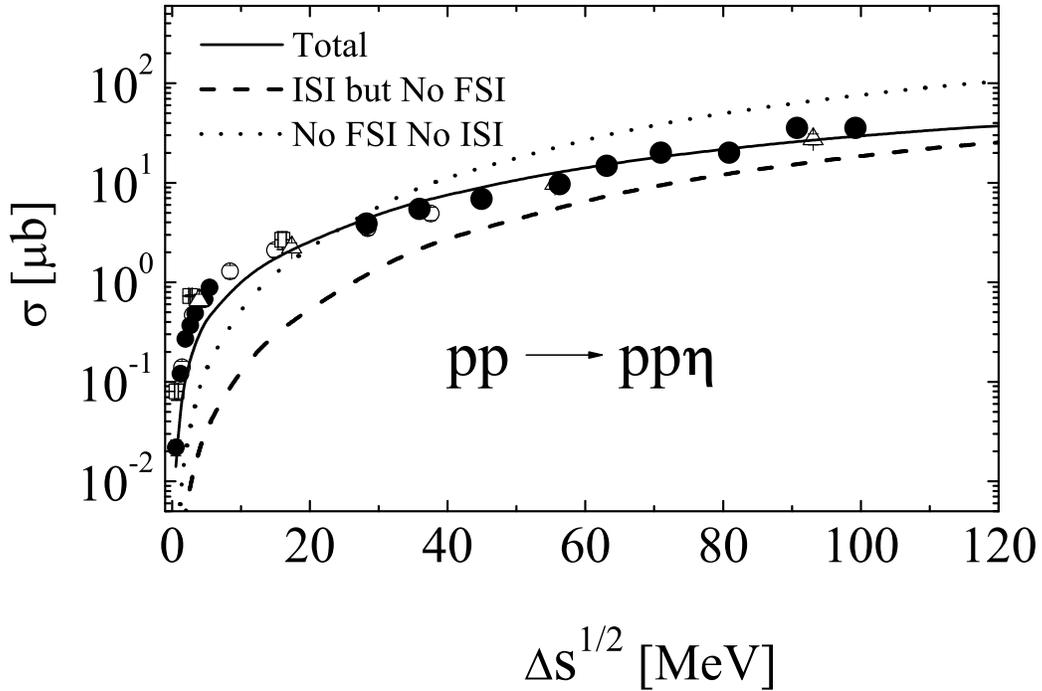} %
\vskip -9mm
\caption{Effect of ISI and FSI for the reaction $pp \to pp \eta$.
The dashed curve depicts the cross section when switching off the FSI,
while the dotted curve is without both FSI and ISI.
The solid curve and data are as in Fig.~\ref{fig3}.}
\label{fig5A}
\end{figure}

\begin{figure}[h]  
\includegraphics[width=1.0\textwidth]{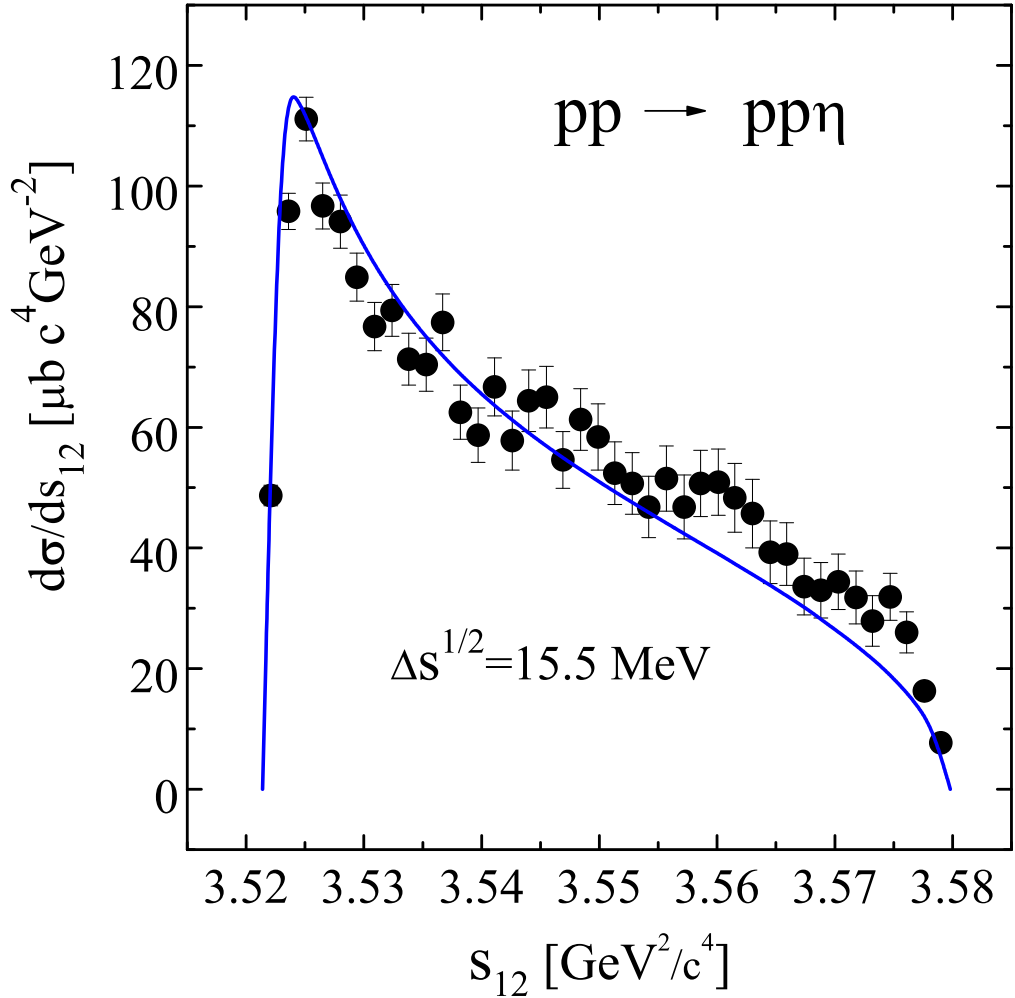} %
\vskip -9mm
\caption{Distribution of the proton-proton invariant mass
$s_{12}$ for the reaction $pp\to pp\eta$ at excess energy
$\Delta s^{1/2}=15.5 \,MeV$. 
Since there is an overall normalization uncertainty of data at
$\Delta s^{1/2}=15.5 - 16.0 \,MeV$ obtained by different groups
our curve has been normalized to the data 
(see Ref.~\cite{etacalen,etaexper3}, normalization factor $\approx 1.4$).
Data are from Ref.~\cite{etaexper3}.}
\label{fig6}
\end{figure}

As mentioned in the introduction, in heavy-ion collisions a substantial part of
di-electrons stem from Dalitz decays of $\eta$ mesons. In the same invariant mass
region also $\Delta$ Dalitz and bremsstrahlung processes contribute to the total
yield. It is, therefore, important to have some reliable cross sections for elementary
$\eta$ production in $pp$ and $pn$ collisions as input for transport model simulations.
Here one faces the problem whether the free cross sections (say, simply parameterizations
of data) should be utilized or such ones where the FSI is ''switched off'', as in a dense
hadronic environment the outgoing nucleons are not asymptotically free out-states.
The same holds for the ISI. To have some indicator of the size of ISI and FSI effects
we show in Fig.~\ref{fig5A} the same as in Fig.~\ref{fig3} but with FSI switched off
(dashed curve) or both ISI and FSI switched off (dotted curve, which may be called
pure production cross section). A similar pattern
holds for $pn \to pn \eta$ (not displayed). One observes indeed sizeable effects
of ISI and FSI which point to the importance of a proper treatment of such effects
in accurate many-body simulations.

\section{The Complete Reaction $\bf pp \to pp \boldeta \to pp \boldgamma e^+e^-$}

In order to emphasize the dependence on the di-electron invariant mass we
rewrite the integrated cross section (\ref{two}) 
$d\sigma / ds_{\gamma^*} = \int ds_\eta \,(d\sigma / ds_\eta ds_{\gamma^*})$
in the form
\begin{eqnarray}&&
\frac{d\sigma}{ds_{\gamma^*}}=\frac{2\alpha}{3\pi^2}
|F_{\eta \gamma \gamma^*} (s_{\gamma^*})|^2
\frac{Br(\eta\to 2\gamma)}{m_\eta^2}{\cal I}(s_{\gamma^*}), \label{xi1} \\&&
{\cal I}(s_{\gamma^*})\equiv\int\limits_1^{\xi_{max}}
d \xi \frac{(\xi-1)^3}{\xi \, a}
\frac{b}{(\xi-a)^2+b^2}
\sigma^{tot}_{NN\to NN\eta}(\xi),
\label{xi}
\end{eqnarray}
where we introduced dimensionless variables $a$, $b$ and $\xi$
\begin{equation}
a\equiv \frac{m_\eta^2}{s_{\gamma^*}};\qquad b\equiv \frac{m_\eta\, \Gamma_\eta}{s_{\gamma^*}},
\qquad \xi\equiv \frac{s_\eta}{s_{\gamma^*}}, 
\quad \xi_{max}=\frac{(\sqrt{s}-2m_\eta)^2}{s_{\gamma^*}}.
\label{pae}
\end{equation}
$d \sigma / d s_{\gamma^*}$ may be considered as normalized contribution
to the di-electron invariant mass spectrum mediated by the $\eta$ Dalitz decay.  

Since the total width of the $\eta$ meson
is fairly small,  
the parameter $b$ in (\ref{xi}) provides a very sharp maximum of the integrand function
at $\xi=a$ as long as the parameter obeys $a>1$ 
(or, equivalently, the di-electron invariant mass fulfills
$s_\gamma^* < m_\eta^2$). This allows to pull out the smooth
function $\sigma^{tot}_{NN\to NN\eta}(\xi=a)$ from the integral and to calculate it at
$s_\eta=m_\eta^2$, i.e. for the on-shell $\eta$ meson. The remaining part can
be calculated analytically. However, for $a < 1$, when the di-electron mass is larger
than the $\eta$ pole mass, the integrand does not exhibit anymore a resonant shape and
the integral ought to be calculated numerically.

Equation (\ref{xi1}) shows that the di-electron invariant mass distribution
$d\sigma / ds_{\gamma^*}$ is  proportional to the transition form factor
$F_{\eta\gamma\gamma^*}(s_{\gamma^*})$ so that measurements of this distribution
provide direct experimental information about the FF.
It can be checked that
up to $a\sim 1.035$ (corresponding to $s_{\gamma^*} \simeq 0.29 \,GeV^2$) the
smooth part of the integrand, $\dfrac{b(\xi-1)^3}{a\xi}\,\sigma^{NN\to NN\eta}(\xi)$, can
be pulled out from the integral at $\xi=a$ obtaining
\begin{eqnarray}
\label{approxint}
{\cal I}(s_{\gamma^*})=
\dfrac{\pi(a-1)^3}{a^2}\,\sigma^{NN\to NN\eta} (s_\eta=m_\eta^2) .
\end{eqnarray}
Equations (\ref{xi1} - \ref{approxint}) allow then to define an experimentally
measurable ratio which is directly proportional to the form factor squared
\begin{eqnarray}
\label{epsFF}
R(s_{\gamma^*})=
\frac{d\sigma/ds_{\gamma^*}}{\left.
\left (d\sigma/ds_{\gamma^*}\right)\right|_{s_{\gamma^*}=s_{\gamma^*, min}}}
\frac{s_{\gamma^*} }{s_{\gamma^*, min}}
\left( \frac{1-s_{\gamma^*, min}/m_\eta^2}{1-s_{\gamma^*}/m_\eta^2}\right)^3=
\frac{|F_{\eta\gamma\gamma^*}( s_{\gamma^*})|^2}{|F_{\eta\gamma\gamma^*}( s_{\gamma^* , min})|^2}
\end{eqnarray}
where $s_{\gamma^* ,min} $ is the minimum value accessible experimentally
(in the ideal case  this is the kinematical limit  $s_{\gamma^* ,min}=4\mu_e^2 $
with electron mass $\mu_e$). At low
enough values of $s_{\gamma^* ,min} $, the transition FF is close to its normalization
point $F_{\eta\gamma\gamma^*}(0)=1$ and the ratio (\ref{epsFF}) is just the transition FF as
a function of $s_{\gamma^*}$.

As $a$ approaches unity keeping  the maximum position  still within the integration range, 
one can
again withdraw from the integral the smooth function 
$\sigma^{NN\to NN\eta}$ at $s_\eta=m_\eta^2$.
However, now the function $(\xi-1)^3/\xi$ can not be considered as smooth enough
and must be kept under the integration. Nevertheless, even
in this case the remaining integral can be computed analytically
(see Appendix B) and one can still define a ratio analogous to (\ref{epsFF}) 
which allows for an experimental
investigation of the FF near the free $\eta$ threshold ($s_{\gamma^*}\to m_\eta^2$).
At $s_{\gamma^*}> m_\eta^2$ the integral ${\cal I}(s_{\gamma^*})$ 
does not have anymore a sharp
maximum and it must be calculated numerically as mentioned above.

\begin{figure}[h]  
\includegraphics[width=1.0\textwidth]{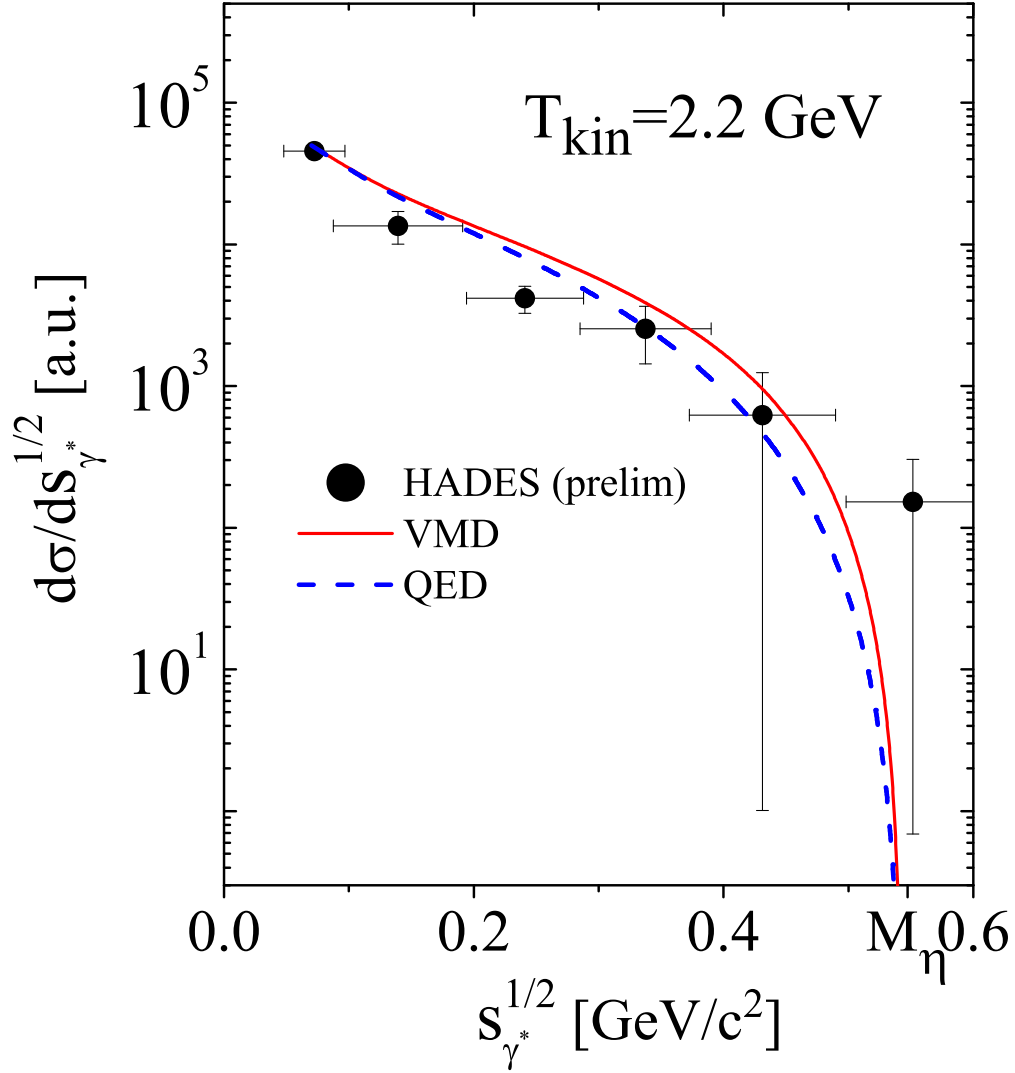} %
\vskip -9mm
\caption{Invariant mass distribution
of di-electrons produced by Dalitz decay of $\eta$ mesons in
the reaction $pp\to pp \gamma e^+e^-$ at initial
kinetic energy $T_{kin}=2.2 \,GeV$.
The solid line is for results with
the $\eta\gamma\gamma^*$ transition FF computed within the VMD model,
while the dashed line corresponds to a pure QED calculation of the
$\eta\gamma\gamma^*$ vertex, i.e., for $\vert F_{\eta \gamma \gamma^*} \vert^2 = 1$.
The position of $\eta$ pole mass is exposed. 
Preliminary experimental data are from Ref.~\cite{hadeseta}.}
\label{fig7}
\end{figure}

In Fig.~\ref{fig7} the di-electron invariant mass distribution 
$d\sigma / d s_{\gamma^*}^{1/2}$
is exhibited as a function of the invariant mass $s_{\gamma^*}^{1/2}$ calculated from
formula (\ref{xi1}). The solid curve employs
the transition FF $F_{\eta\gamma\gamma^*}(s_{\gamma^*})$ calculated
within the VMD model \cite{LandPhysRep}
\begin{equation}
F_{\eta\gamma\gamma^*}^{VMD}(s_{\gamma^*}) =
\dfrac{m_\rho^2}{s_{\gamma^*} -(m_\rho- \frac{i}{2} \Gamma_\rho)^2}
\label{VMD_FF}
\end{equation}
with standard values for $m_\rho$ and $\Gamma_\rho$ \cite{dataGroup}.
(Note that in the kinematical region of interest the $\rho$ contribution
is sufficient.)
The dashed curve in Fig.~\ref{fig7} is for the pure QED part, i.e.\ without accounting for the
strong form factor, i.e.\ with $F_{\eta\gamma\gamma^*}(s_{\gamma^*})=1$.
The experimental data
in Fig.~\ref{fig7} are preliminary results
from HADES \cite{hadeseta}. It is clearly seen that without FF the calculated cross section
rapidly drops as $s_{\gamma^*}^{1/2}\to m_\eta$, while inclusion of the FF 
only mildly modifies the shape
(factor 2 increase at $m_{\gamma^*} \sim 400 \,MeV$ 
High precision data are needed to arrive at a firm conclusion on the
validity of the employed VMD FF.

\begin{figure}[h]  
\includegraphics[width=1.0\textwidth]{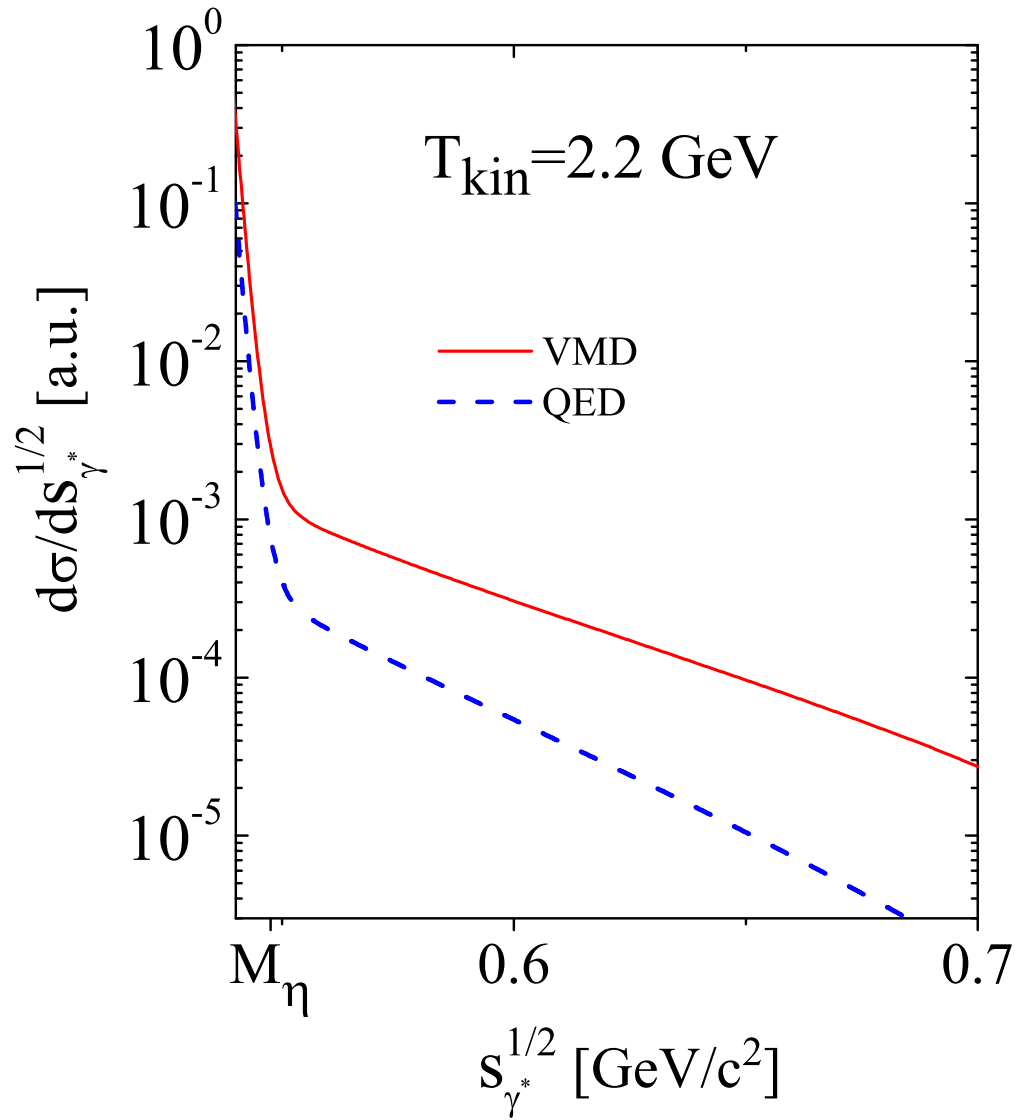} %
\vskip -9mm
\caption{ The same as in Fig.~\ref{fig7} but
in the kinematical region beyond the "$\eta$ threshold", i.e.
for di-electron masses larger than the $\eta$ pole mass
being accessible for intermediate off-shell $\eta$ mesons.}
\label{fig8}
\end{figure}

With increasing $s_{\gamma^*}^{1/2}$, the effect of
the FF becomes more and more pronounced. This can be visualized if one calculates the
$e^+ e^-$ mass distribution beyond the free $\eta$ meson pole mass, i.e.
at $s_{\gamma^*}^{1/2}>m_\eta$. Figure \ref{fig8} exhibits the behavior
of the mass distribution at large values of $s_{\gamma^*}^{1/2}$.
The effect of the FF increases noticeably at large values of $s_{\gamma^*}$. 
This is of interest since, as mentioned, the VMD calculations provides a good description of
the on-shell $\eta$ meson FF in the range $s_{\gamma^*}<m_\eta^2$.
An investigation of the FF
at $s_{\gamma^*} > m_\eta^2$ can provide information of the relevance of the VMD
model for off-shell hadrons at kinematical limit.
In particular, when calculating Dalitz type processes with (off-shell) vector mesons
created via bremsstrahlung off nucleon resonances,
one usually estimates the couplings in the corresponding vertices from the
radiative decay off an on-shell resonance with applying the VMD model.
An additional justification of such a procedure can be obtained
from investigating the transition FF beyond the on-mass shell limit.
It should be noted, however, that severe background processes will make difficult
an identification of the $\eta$ Dalitz yield in this kinematical region.

\section{Summary}\label{sumary}

In summary we have analyzed the di-electron production from Dalitz decays
of $\eta$ mesons produced in $pp$ collisions at intermediate energies.
The corresponding cross section has been
calculated within an effective meson-nucleon approach with
parameters adjusted to a large extent to describe the free vector meson production 
in $NN$ reactions near the threshold. We argue that by studying
the invariant mass distribution of  the final $e^+e^-$ system
in a large  kinematical interval of 
di-electron  masses, one can directly measure
the $\eta$ meson transition form factor $F_{\eta \gamma \gamma^*}$ in, e.g., $pp$
collisions. Such experiments are already performed and will be further improved
at HADES. Our results may serve as prediction for these
forthcoming experiments. The uncertainties of such a procedure
depend upon the scale of the background processes and is expected to be
small if the interference is destructive \cite{ourFF}. Experimental information
on transition form factors is useful for testing QCD predictions of hadronic quantities
in the non-perturbative domain.

\section*{Acknowledgements}
We thank  H.W. Barz and A.I. Titov for useful discussions.
L.P.K. would like
to thank for the warm hospitality in the Research Center Dresden-Rossendorf.
This work has been supported by
BMBF grants 06DR121, 06DR136, GSI and the Heisenberg-Landau program.

\appendix

\section{Coupling constants for spin-$\frac32$ resonances}

The coupling constants of the spin-$\frac32$ resonance $D_{13}(1520)$ can be estimated
from the known decay widths. For instance, for  the decay $D_{13}(1520)\to N \eta$
one has
\begin{eqnarray}
\label{b1}
\Gamma_{N^*\to N\eta} &=&
\frac{1}{8\pi}\frac{\sqrt{[m_{N^*}^2 -(m_N-m_\eta)^2][m_{N^*}^2 -(m_N+m_\eta)^2]}}{2m_{N^*}^3}
\frac{g^2_{\eta NN^*}}{m_\eta^2}\frac14 \sum_{spins} | {\cal M}|^2\nonumber\\
&=&
\frac{1}{64\pi m_{N^*}^3 }\sqrt{[m_{N^*}^2 -(m_N-m_\eta)^2][m_{N^*}^2 -(m_N+m_\eta)^2]}
\frac{g^2_{\eta NN^*}}{m_\eta^2} \frac23 \left(\frac{(P_{N^*} q_\eta)}{m_{N^*}^2}-m_\eta^2
\right)\nonumber\\
&\times&
{\rm Tr} \left[(\hat P_N +m_N)\gamma_5(\hat P_{N^*}+m_{N^*})\gamma_5) \right],
\end{eqnarray}
where for the spin-$\frac32$ particles we use the Rarita-Schwinger spinors
${\cal U}^\mu (P_{N^*}, \bar s )$ for the spin $\bar s$ summation 
with the relation
\begin{eqnarray}
\label{r1}
\sum_{\bar s} {\cal U}^\mu(P_{N^*}, \bar s) \bar {\cal U}^\nu(P_{N^*}, \bar s) &=&
(\hat P_{N^*}+m_{N^*})\nonumber \\
&\times&
\left( -g_{\mu\nu} +\frac{ \gamma^\mu\gamma^\nu}{3} +
\frac{(\gamma^\mu  P_{N^*}^\nu-\gamma^\nu  P_{N^*}^\mu)}{3m_{N^*}}
+ \frac{2 P_{N^*}^\mu P_{N^*}^\nu }{3m_{N^*}^2} \right).
\end{eqnarray}
We use the short hand notation $\hat p = (p \cdot \gamma)$.
Equation (\ref{b1}) provides
slightly larger coupling constants than that used in Ref.~\cite{nakayamaEta}
in which, besides the decay widths, the coupling constants have been adjusted to fit also other
processes involving $D_{13}(1520)$. 
An analog equation hold for $D_{13}(1520)\to N \pi$.
In the present paper we take $g_{\pi NN^*}=1.55$ and
$g_{\eta NN^*}=8.3$, see Tab.~\ref{tb2}.

\section{Integration over the variable $\bf s_\boldeta$  }\label{appA}

Since the parameter $b$ in eq.~(\ref{xi}) is very small
in the whole kinematical region of $s_{\gamma^*}$ the
integrand has a sharp maximum around $\xi=a$. If $a>x_{min}=1$ then the
maximum is located inside the integration interval of $\xi$ and one can
take advantage of the smoothness of the $\eta$ production cross section
$\sigma^{NN\to NN\eta}$ to withdraw it from the integral. Then
\begin{eqnarray}
\label{appen1}
{\cal I}_1(\xi)&\equiv &\int d\xi \frac{(\xi-a)^3}{\xi}\frac{b/a}{(\xi-a)^2 +b^2}\nonumber\\
&=&
\frac{b}{a}\xi +\frac{b}{a(a^2+b^2)}\left[
-\ln \xi +\frac12 \ln\left( (\xi-a)^2 +b^2\right)\left (
2a^3 -3a^2 + 2ab^2 -3b^2+1\right) \right. \nonumber\\
&+& \left.
\frac{1}{b}{\rm arctan}\left(\frac{\xi-a}{b}\right)
\left( a^4 -3a^3 +3a^2  -a -3ab^2 -b^4+3b^2\right) \right] .
\end{eqnarray}
It is seen that in case when $a\gg 1$, $b\ll 1$ the term $b \xi / a$ and terms
containing logarithms in (\ref{appen1}) can be disregarded:
\begin{eqnarray}
\label{app2}
{\cal I}_1(\xi) &\simeq &\frac{(a-1)^3}{a^2} {\rm arctan}\left(\frac{\xi-a}{b}\right),\\
\label{app3}
{\cal I}_1(\xi_{max}) -{\cal I}_1(\xi_{min}) &\simeq &
\pi\frac{(a-1)^3}{a^2}.
\end{eqnarray}
This corresponds to the case when in eq.~(\ref{xi}) one withdraws from the integral
both the production cross section
$\sigma^{NN\to NN\eta}$ and the combination $(\xi -1)^3 / \xi$ at $\xi=a$. Note that
eq.~(\ref{appen1}) is a good approximation even for $a<1$, if $a+b >1$, i.e. for $a$ in the
$b$ vicinity of unity.

\newpage

\newpage

\end{document}